\documentclass[runningheads]{llncs}
\usepackage{graphicx}
\usepackage{bbding}
\usepackage{graphicx}
\usepackage{float}
\usepackage{subfigure}
\usepackage{booktabs}
\usepackage{makecell}
\usepackage{multirow}
\usepackage{caption}
\usepackage{amsmath}
\usepackage[table]{xcolor}
\usepackage[colorlinks=true, citecolor=blue, linkcolor=blue, urlcolor=blue]{hyperref}

\begin{document}
\title{DA-Lion: Efficient Neural Video Representation via Direction-Aware Optimization}
\titlerunning{Efficient NVR via Direction-Aware Optimization}

% \author{Anonymous author}
\author{Qingyu Mao\inst{1} \and
Jiacong Chen\inst{2,3} \and
Shuai Liu\inst{2,3} \and
Yongsheng Liang\inst{1,3} \and
Youneng Bao\inst{1,(} \Envelope \inst{)}
}

\authorrunning{Q. Mao et al.}

\institute{College of Electronics and Information Engineering, Shenzhen University \and
College of Applied Technology, Shenzhen University \and
School of Artificial Intelligence, Shenzhen Technology University \\
% Research Center of Networks and Communications, Peng Cheng Laboratory, Shenzhen, China \and
% Faculty of Artificial Intelligence in Education, Central China Normal University \\
\email{baoyn@szu.edu.cn} }
\maketitle              % typeset the header of the contribution
\begin{abstract}
Implicit neural video representation (NVR) encodes a video as the parameters of an overfitted neural network, where training and testing data are identical and the objective is instance-specific signal fitting.
In this deterministic regime, optimization dynamics directly determine reconstruction quality under a fixed budget, yet existing methods universally adopt general-purpose optimizers designed for stochastic gradient training.
Adam-family optimizers descend steadily but converge slowly, while Lion achieves faster early progress but oscillates markedly in later stages when gradient directions become unstable.
We trace this instability to the insensitivity of the sign-based update to gradient--momentum alignment.
To address this, we propose Direction-Aware Lion (DA-Lion), a task-driven optimizer tailored for NVR.
DA-Lion introduces (i) a direction-consistency criterion that switches between sign update and momentum update based on gradient--momentum alignment, and (ii) a learning-rate-aware magnitude modulation that stabilizes effective step sizes across training phases. 
DA-Lion modifies only the parameter update rule, preserving model architecture, parameter count, FLOPs, and decoding throughput.
Experiments on three datasets with four NeRV-family backbones show that DA-Lion improves convergence stability and boosts PSNR/MS-SSIM across all backbones.
The source code will be made available at \url{https://github.com/maoqingyu1996/DA-Lion}.

\end{abstract}

\section{Introduction}
Implicit neural representations (INRs)~\cite{sitzmann2020siren,mildenhall2021nerf} have emerged as an effective alternative for visual data compression~\cite{mao2025wdsinr}, representing a target signal by the parameters of a neural network that maps coordinates (e.g., pixel positions) to signal values (e.g., RGB colors). For compression, this network is overfitted to a single instance---the learned parameters serve as a compact code for reconstruction via a forward pass.

In neural video compression, NeRV~\cite{chen2021nerv} and its variants~\cite{li2022e-nerv,chen2023hnerv,zhao2024pnerv,chen2025lthnvr} represent a video as a compact network $f_\theta$ that maps a temporal embedding $e_t$ to its corresponding RGB frame $I_t$.
Despite their architectural diversity, all NeRV-family methods share the same encoding pipeline: (i) define a network architecture $\mathcal{A}$, (ii) set the initial parameters $\theta_{\mathcal{A}}^{0}$ and temporal embeddings $e_t^0$, and (iii) train with loss $\mathcal{L}$ and optimizer $\mathrm{O}$ to obtain the optimized parameters $\theta_{\mathcal{A}}^{*}$ and embeddings $e_t^*$:
\begin{equation}
\theta_{\mathcal{A}}^{0}, e_t^0 \xrightarrow[\text{set}]{} \mathcal{L},\, \mathrm{O} \xrightarrow[\text{train}]{} \theta_{\mathcal{A}}^{*},\, e_t^*,
\label{eq:pipeline}
\end{equation}
where $\mathrm{O}$ drives the search from $\theta_{\mathcal{A}}^0$ to $\theta_{\mathcal{A}}^*$ by iteratively minimizing $\mathcal{L}$. The trained parameters $\theta_{\mathcal{A}}^*$ (with $e_t^*$) are stored as the compressed bitstream, and decoding evaluates $f_{\theta_{\mathcal{A}}^{*}}(e_t^*)$ for each frame.

\begin{figure}[t]
\centering
\includegraphics[width=\textwidth]{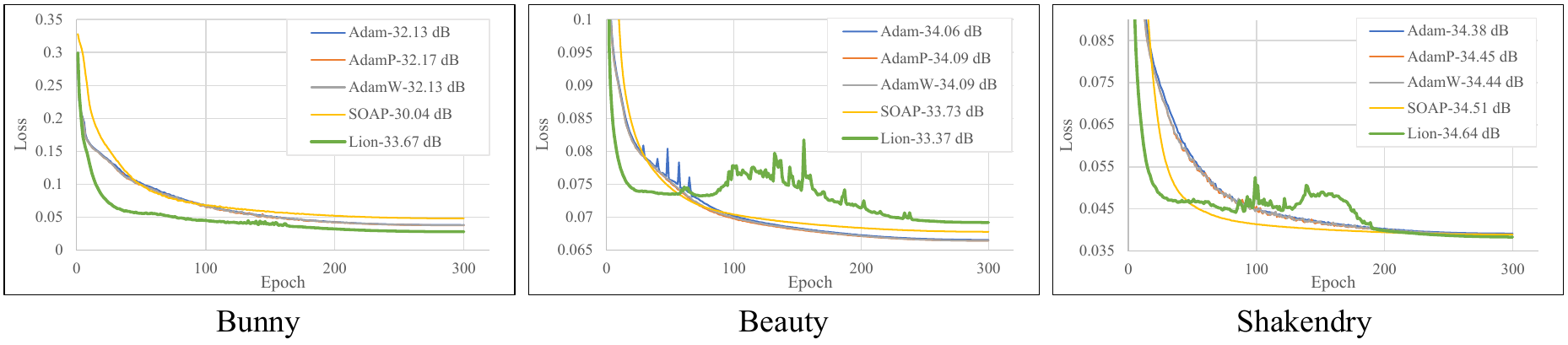}
\caption{Training loss curves of five general-purpose optimizers on three video sequences. Adam-family optimizers (Adam, AdamW and AdamP) and SOAP descend steadily but converge slowly with a higher final loss, while Lion converges rapidly at early stages but tends to oscillate, leading to suboptimal final performance.}
\label{fig1}
\vspace{-1.0em}
\end{figure}

Existing efforts on NeRV-family methods focus on architectural~\cite{zhao2023dnerv,kwan2024hinerv,li2025onlinerepnerv} or objective function~\cite{kim2024snerv,tang2023scene} innovations, while universally adopting Adam~\cite{kingma2015adam} or AdamW~\cite{loshchilov2017adamw} as the default optimizer. 
The choice of optimizer, however, has received surprisingly little scrutiny.
Standard optimizers are designed for mini-batch training with inherent gradient noise.
In NVR, training is deterministic and instance-specific: gradients reflect the evolving fitting state rather than sampling noise, and the objective is to minimize distortion on the identical coordinate set used for evaluation.
Late-stage refinement requires stable, high-precision updates robust to directional fluctuations---not mechanisms compensating for stochastic gradient noise.
In the absence of gradient noise, Adam's adaptive rates become unnecessarily conservative; conversely, Lion's sign-based updates discard magnitude information, making the update vulnerable to directional instability where small gradient fluctuations can trigger full-magnitude sign reversals.
Fig.~\ref{fig1} confirms these patterns: Adam-family optimizers descend steadily but converge slowly; Lion accelerates early progress but oscillates in later stages.

In this work, we propose Direction-Aware Lion (DA-Lion), a task-driven optimizer tailored for NVR. Building on our analysis of Lion's instability---where sign-based updates amplify directional fluctuations when the gradient opposes historical momentum (see Sec.~\ref{sec:method})---DA-Lion introduces two complementary components: a \textbf{direction-consistency criterion} that switches between sign update and momentum update based on gradient--momentum alignment, and a \textbf{learning-rate-aware magnitude modulation} that stabilizes effective step sizes across training phases. 
DA-Lion modifies only the parameter update rule, preserving model architecture, parameter count, FLOPs, and decoding throughput.
Experiments across four NeRV-family backbones and three video benchmarks show that DA-Lion improves convergence stability and reconstruction quality.

\section{Related Work}
\noindent\textbf{Implicit neural video representation.}
NVR represents a video as a compact network optimized per video. NeRV~\cite{chen2021nerv} maps frame indices to RGB frames; subsequent work improves reconstruction via architectural refinements (E-NeRV~\cite{li2022e-nerv}, HNeRV~\cite{chen2023hnerv}, D-NeRV~\cite{he2023d-nerv}, HiNeRV~\cite{kwan2024hinerv}, DNeRV~\cite{zhao2023dnerv}, PNeRV~\cite{zhao2024pnerv}), meta-learned initialization (MetaNeRV~\cite{guo2025metanerv}), or temporally coherent modulation (NVTM~\cite{shin2024nvtm}). Practical codecs study end-to-end compression pipelines~\cite{zhang2021vcompress,gao2025pnvc}. These directions are complementary: we improve performance by adapting the optimizer without modifying the backbone.

\noindent\textbf{Optimization for deep networks.}
Adam~\cite{kingma2015adam} and AdamW~\cite{loshchilov2017adamw} stabilize training under stochastic gradients; Lion~\cite{chen2023lion} uses sign-based updates on a momentum direction, speeding up early convergence by discarding gradient magnitudes.
A related line of work conditions the update on gradient--momentum discrepancy.
AdaBelief~\cite{zhuang2020adabelief} replaces Adam's second moment with the centered variance $(g_t-m_t)^2$, shrinking step sizes when gradient and momentum disagree.
Cautious Optimizers~\cite{liang2024cautious} mask out coordinates where $\mathrm{sign}(g_t)$ disagrees with momentum, leaving them unchanged.
DA-Lion differs in two ways: (i) instead of masking disagreeing coordinates to zero, it switches them to a momentum-form update, preserving descent rather than freezing; (ii) it pairs the direction-consistency criterion with a learning-rate-aware magnitude modulation that further stabilizes step sizes during late refinement.
In NVR, gradients reflect the evolving fitting state rather than mini-batch noise, making late-stage refinement sensitive to directional fluctuations---motivating update rules that explicitly respond to direction instability.

\section{Method}\label{sec:method}
\subsection{Revisiting Lion for NVR}
\begin{figure}[htbp]
\centering
\vspace{-1.5em}
\includegraphics[width=0.95\textwidth]{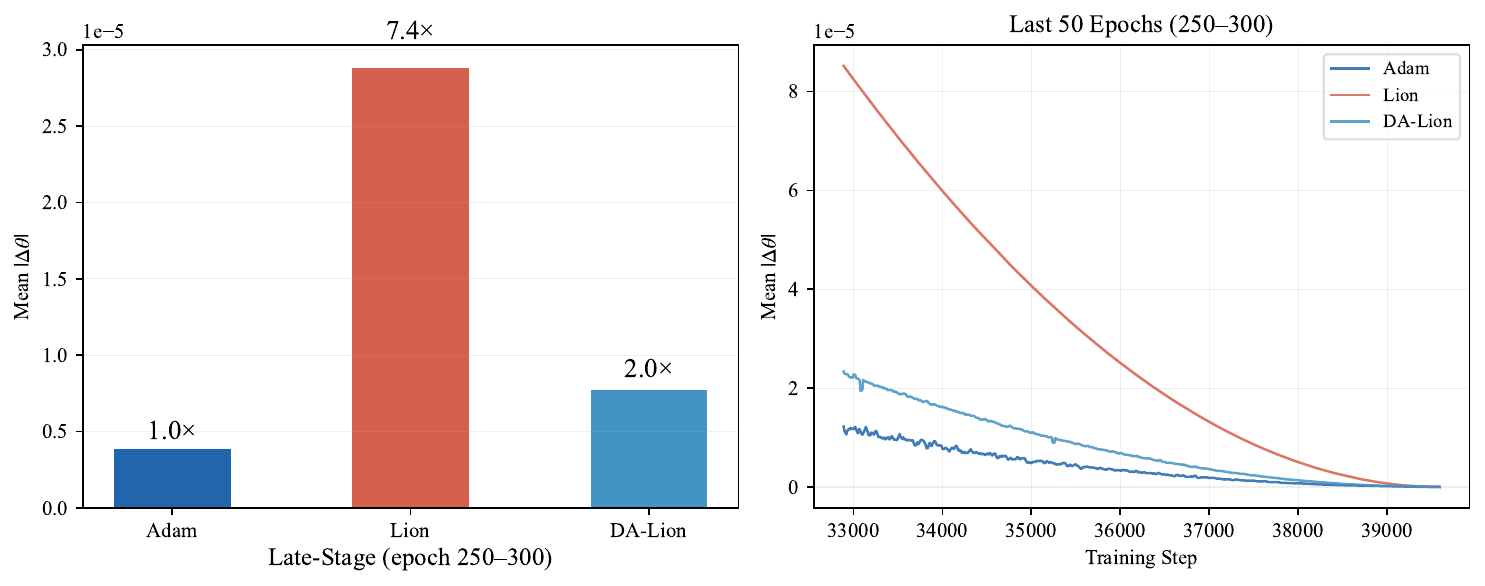}
\caption{Quantitative noise-floor comparison. Left: mean $|\Delta\theta|$ over the final 50 epochs. Right: per-step trajectory during the last 50 epochs. Lion's update magnitude is structurally decoupled from the gradient, tracking the cosine LR schedule with negligible variation ($0.4\%$); Adam ($7\%$) and DA-Lion ($1.9\%$) exhibit content-driven fluctuations.}
\label{fig:noise_quant}
\vspace{-1.0em}
\end{figure}

We first review Lion~\cite{chen2023lion}. Let $\theta_t$ denote parameters and $g_t=\nabla_{\theta} L(\theta_{t-1})$ the gradient at iteration $t$.
Lion maintains an exponential moving average (EMA) of historical gradients, updated by
\begin{equation}
m_t = \beta_2 m_{t-1} + (1-\beta_2) g_t.
\label{eq:lion_mom}
\end{equation}
At each step, Lion computes a momentum-like direction
\begin{equation}
u_t = \mathrm{sign}\!\left(\beta_1 m_{t-1} + (1-\beta_1) g_t\right),
\label{eq:lion_dir}
\end{equation}
then updates $\theta_t = \theta_{t-1} - \eta_t\, u_t$ with learning rate $\eta_t$.
This sign-based acceleration, however, introduces a structural failure mode in deterministic NVR.
Because $\mathrm{sign}(\cdot)$ is discontinuous at zero, an arbitrarily small fluctuation of $\tilde m_{t,i}$ across zero is amplified into a full $\pm\eta_t$ step in coordinate $i$, yielding $|\Delta\theta_{t,i}|=\eta_t$ irrespective of the gradient magnitude.
Near a local minimum, gradients are dominated by inter-coordinate residual error rather than a coherent descent direction; the per-parameter trajectory is driven by a ``noise-floor'' jitter of magnitude $\eta_t$.
Cosine annealing shrinks $\eta_t$ but cannot eliminate this jitter, and unlike stochastic training, NVR provides no mini-batch averaging to suppress it.
Fig.~\ref{fig:noise_quant} provides direct quantitative evidence: at the late training stage, Lion's mean per-parameter update magnitude is $7.4\times$ that of Adam, confirming the noise floor maintains elevated update amplitudes throughout refinement.

To address this, we propose Direction-Aware Lion (DA-Lion), which selectively suspends sign-based updates on coordinates exhibiting noise-driven jitter while preserving Lion's fast progress on coordinates in stable descent.

\subsection{Direction-aware update}
DA-Lion retains the momentum construction of Eq.~(\ref{eq:lion_mom}) but uses a direction-aware update rule.
We define a per-parameter direction-consistency score between the current gradient and the historical momentum as the element-wise product:
\begin{equation}
\Delta_t = g_t \odot m_{t-1},
\label{eq:consistency}
\end{equation}
where each entry $\Delta_{t,i}=g_{t,i}\,m_{t-1,i}$ independently judges the alignment of coordinate~$i$.
Using $m_{t-1}$ rather than $m_t$ or $\tilde m_t$ is intentional: $m_{t-1}$ reflects the accumulated direction \emph{before} observing $g_t$, so $\Delta_{t,i}>0$ means the new gradient agrees with the historical trend. In contrast, both $m_t$ and $\tilde m_t$ already contain $g_t$, which would make the indicator self-referential and systematically overestimate alignment near convergence.
The threshold at zero requires no task-specific calibration: $\Delta_{t,i}>0$ is equivalent to $\mathrm{sign}(g_{t,i})=\mathrm{sign}(m_{t-1,i})$, a minimal geometric condition for directional agreement.

For each coordinate, DA-Lion applies a per-element switching rule.
Let $\tilde{m}_t=\beta_1 m_{t-1} + (1-\beta_1) g_t$ denote the mixed direction.
When $\Delta_{t,i}>0$, the gradient agrees with historical momentum---the coordinate receives a sign-based update for fast progress.
When $\Delta_{t,i}\le 0$, the gradient opposes the historical direction---the coordinate receives a momentum-form update that preserves gradient magnitude, preventing the noise-floor amplification analyzed in Sec.~\ref{sec:method}.
Formally, with the per-element mask $s_t=\mathbf{1}[\Delta_t>0]$, DA-Lion computes
\begin{equation}
\Delta \theta_t = -\alpha_t \cdot \mathrm{sign}(\tilde{m}_t)\odot s_t \;-\; \tilde{m}_t\odot(\mathbf{1}-s_t),
\label{eq:da_lion_update}
\end{equation}
where $\alpha_t$ is a magnitude modulation factor defined in Sec.~\ref{sec:modulation} and the parameter update is $\theta_t=\theta_{t-1}+\eta_t \Delta\theta_t$.

The two branches operate at different numerical scales: the sign branch produces updates of magnitude $\alpha_t\eta_t$, while the momentum branch produces updates scaled by $|\tilde{m}_{t,i}|$. The per-coordinate switching resolves this naturally---each coordinate selects the branch dictated by its local alignment, and the emergent $p_t$ decline (Sec.~\ref{sec:convergence}) ensures a smooth global transition between the two regimes without explicit scale matching.

\begin{figure}[htbp]
\centering
\vspace{-1.5em}
\subfigure[Backbone comparison (Bunny)]{\includegraphics[width=0.48\textwidth]{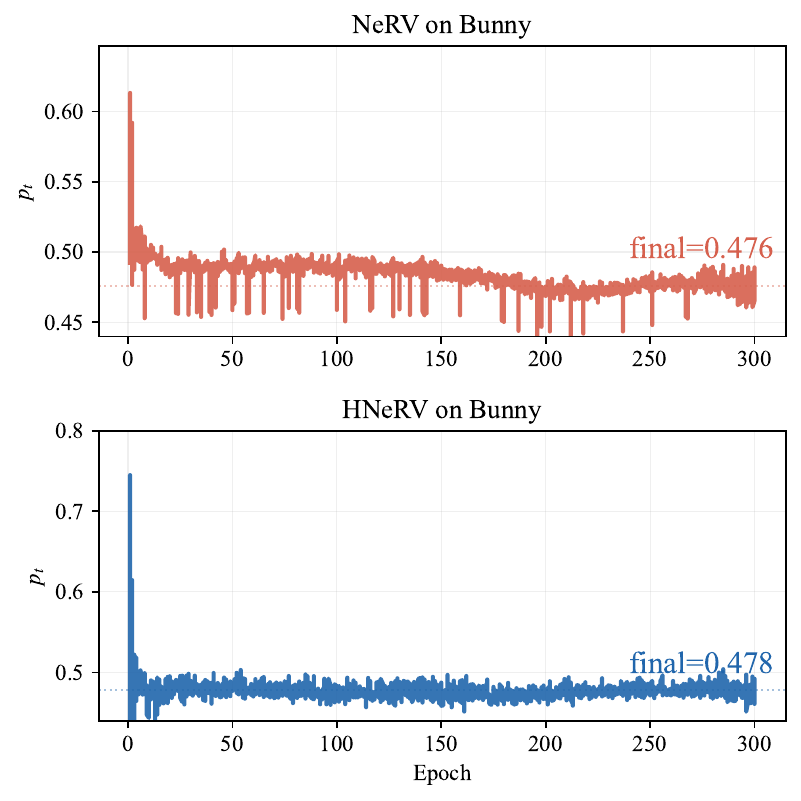}}
\hfill
\subfigure[Video comparison (NeRV)]{\includegraphics[width=0.48\textwidth]{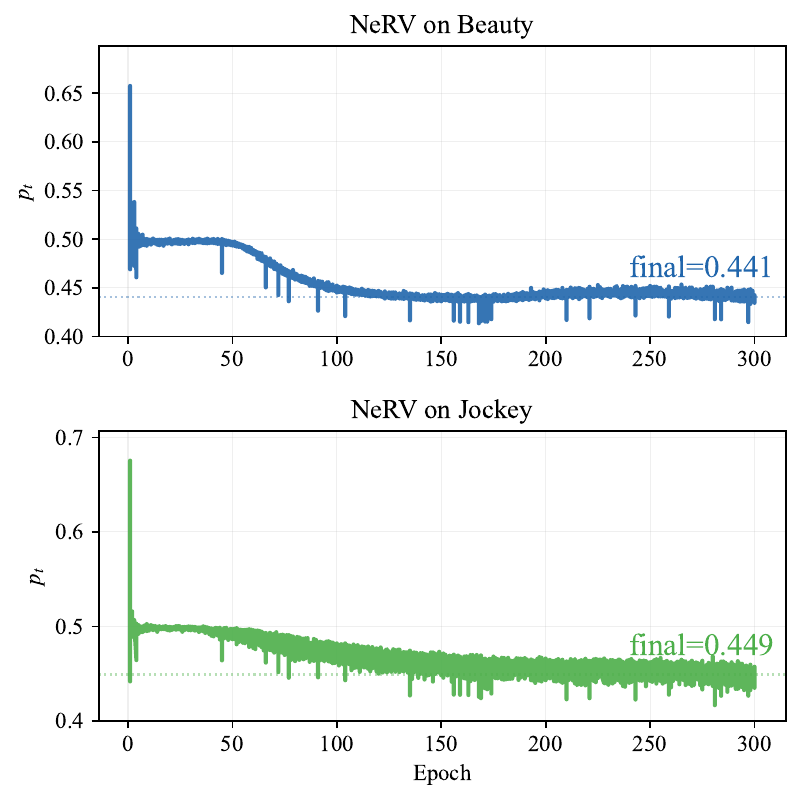}}
\caption{Emergent self-paced phase transition: agreement ratio $p_t$ during DA-Lion training (300 epochs). (a) Bunny: NeRV vs.\ HNeRV. (b) NeRV: Beauty vs.\ Jockey.}
\label{fig:agreement_phase}
\vspace{-1.0em}
\end{figure}

A key property of the per-coordinate criterion is that it induces an implicit, data-driven phase transition.
Define the agreement ratio $p_t=\tfrac{1}{N}\sum_i \mathbf{1}[\Delta_{t,i}>0]$, measuring the fraction of parameters in stable descent.
Fig.~\ref{fig:agreement_phase} reports $p_t$ throughout DA-Lion training.
On Bunny, $p_t$ declines monotonically from $\sim$1.0 to $\sim$0.48 with both NeRV and HNeRV backbones (Fig.~\ref{fig:agreement_phase}a), confirming the transition is governed by the optimizer's internal state rather than architectural specifics.
On the more challenging Beauty and Jockey sequences, $p_t$ reaches $0.44$--$0.45$ (Fig.~\ref{fig:agreement_phase}b), indicating that harder content induces a faster, deeper shift toward momentum-form updates.
All curves decay smoothly without external scheduling---the transition emerges purely from the evolving gradient--momentum alignment, requiring no hand-crafted phase-switching heuristics.

\subsection{Learning-rate-aware magnitude modulation}\label{sec:modulation}
Even with the direction-consistency criterion, the sign branch still applies $\pm\eta_t$ to all aligned coordinates, regardless of the refinement stage.
We introduce a learning-rate-aware modulation factor $\alpha_t$ to decouple the effective sign-step magnitude from $\eta_t$.
Let $r_t = \eta_t/\eta_0$ denote the ratio of the current learning rate to its initial value, which decreases from $1$ to $0$ under standard cosine scheduling.
We define
\begin{equation}
\alpha_t = \alpha_{\min} + (\alpha_{\max}-\alpha_{\min}) \cdot \frac{1}{1+\exp\!\left(s\,(r_t - c)\right)},
\label{eq:alpha}
\end{equation}
with $\alpha_{\min}=0.1$, $\alpha_{\max}=1.0$, slope $s=0.2$, and center $c=0.4$.
The sigmoid form has three deliberate properties.
First, the center $c=0.4$ positions the transition in the mid-to-late training stage where sign-based updates begin to destabilize (cf.\ Fig.~\ref{fig:agreement_phase}), ensuring modulation activates only when needed.
Second, it is strictly bounded in $[0.1,1.0]$, guaranteeing $\eta_t\alpha_t$ never exceeds $\eta_t$; in contrast, a pure exponential is unbounded and exceeds $1.0$ when $r_t<0.05$, amplifying rather than suppressing the step size during late refinement.
Third, the mild slope $s=0.2$ keeps $\alpha_t$ in a narrow band of $[0.52,0.57]$ across all 300 epochs, providing gentle stabilization rather than aggressive step-size reduction.

\begin{figure}[htbp]
\centering
\vspace{-1.5em}
\includegraphics[width=0.95\textwidth]{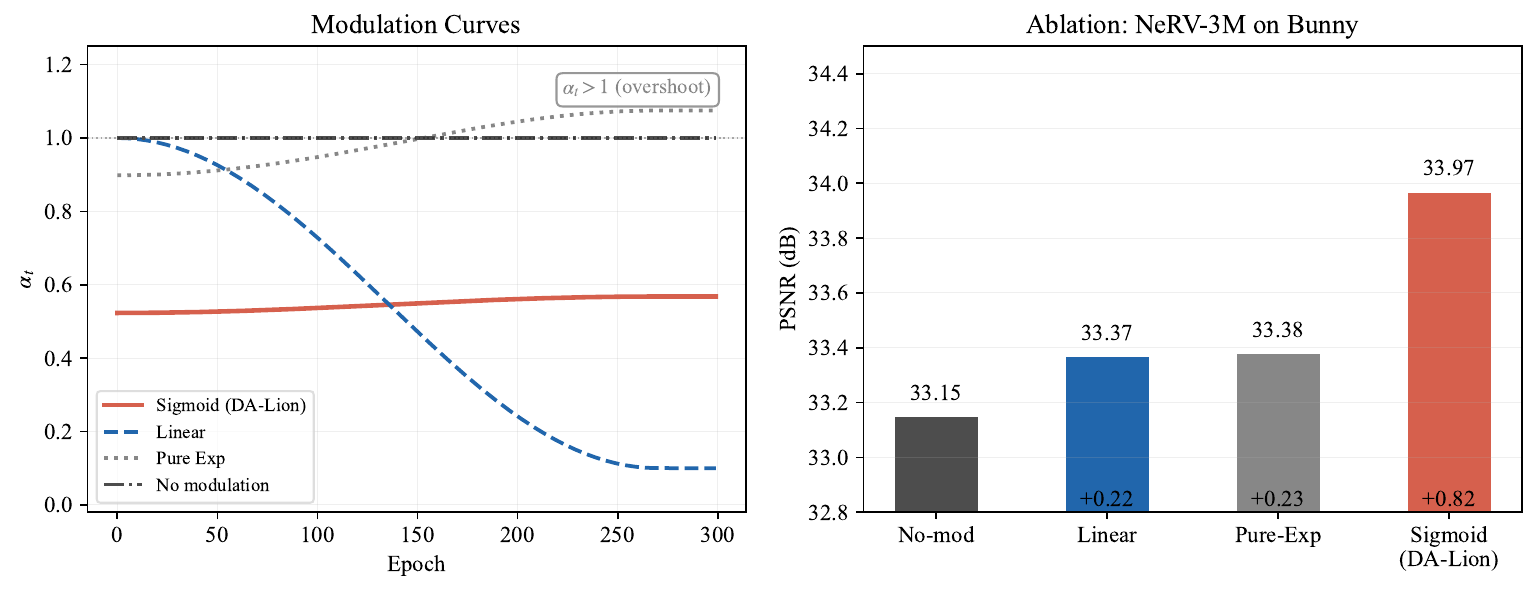}
\caption{Modulation function ablation. Left: $\alpha_t$ curves under a 300-epoch cosine schedule. Right: PSNR on Bunny (NeRV).}
\label{fig:alpha_curves}
\vspace{-1.5em}
\end{figure}

Fig.~\ref{fig:alpha_curves} compares the sigmoid against three alternatives on Bunny.
The sigmoid achieves 33.97\,dB, outperforming the alternatives by $+0.59$--$+0.82$\,dB. The exponential and linear variants perform similarly, confirming that late-stage overshoot cancels any benefit from steeper early decay.

\subsection{Discussion on Convergence}\label{sec:convergence}

We outline a component-wise argument for stability, deferring a rigorous non-convex convergence proof.
When $\Delta_{t,i}>0$, $\tilde m_{t,i}$ is a convex combination of $g_{t,i}$ and $m_{t-1,i}$ sharing their sign; the sign-based update $-\alpha_t\cdot\mathrm{sign}(\tilde m_{t,i})$ is locally anti-aligned with the gradient---a descent step of magnitude $\alpha_t\eta_t$.
When $\Delta_{t,i}\le 0$, the momentum $\tilde m_{t,i}$ carries gradient magnitude that sign updates discard, avoiding the noise-floor amplification identified in Sec.~\ref{sec:method}.
The switching decision depends on $m_{t-1}$, not $m_t$, and is causal.
The agreement ratio $p_t$ decays monotonically from ${\approx}1.0$ to ${\approx}0.44$--$0.48$ (Fig.~\ref{fig:agreement_phase}) without external scheduling, smoothly interpolating from a sign-dominant early phase to momentum-dominant refinement with no abrupt mode switch.
A spurious fixed point would require $\tilde m_{t,i}{=}0$ simultaneously for all disagreeing coordinates, which cannot persist under non-zero residual error.
Per-coordinate step magnitude is bounded by $\eta_t$ ($\alpha_t\in(0,1]$ in the sign branch; $|\tilde m_{t,i}|$ bounded by recent gradient norms otherwise), so $\|\Delta\theta_t\|\to 0$ as the cosine schedule drives $\eta_t\to 0$.
DA-Lion thus interpolates between sign-based exploration and momentum-based refinement without instability, with eventual settling guaranteed by the decaying learning rate.
A formal analysis under standard assumptions ($L$-smoothness, bounded gradients) is left to future work.

\section{Experiments}\label{sec:experiments}

\subsection{Experimental detail}
We evaluated DA-Lion on Bunny, UVG~\cite{mercat2020uvg}, and DAVIS~\cite{perazzi2016davis}. Bunny contains 132 frames at 720$\times$1280; UVG and DAVIS each comprise seven 1080$\times$1920 sequences. All inputs were center-cropped to 640$\times$1280 or 960$\times$1920 for consistent comparison.

We adopted four NVR backbones (NeRV~\cite{chen2021nerv}, E-NeRV~\cite{li2022e-nerv}, HNeRV~\cite{chen2023hnerv}, PNeRV~\cite{zhao2024pnerv}). DA-Lion replaces only the parameter update rule, preserving architectures and compression pipelines. Baselines include Adam~\cite{kingma2015adam} with $(\beta_1,\beta_2)=(0.9,0.999)$ and Lion~\cite{chen2023lion} with $(0.9,0.99)$; DA-Lion inherits Lion's hyperparameters. All models were reproduced from official codebases, trained for 300 epochs at a fixed 3.0M parameter budget on an NVIDIA V100 GPU. We use batch size 1, the Fusion6 loss~\cite{chen2021nerv}, and a cosine learning rate schedule decaying from $10^{-3}$ to $10^{-4}$, following the standard configuration of each backbone.
We report PSNR and MS-SSIM~\cite{wang2003ms-ssim} for reconstruction quality, and model size, bits-per-pixel (bpp), per-iteration training time, encoding time, and GPU memory for complexity.

\subsection{Convergence and Update Dynamics}\label{sec:update_dynamics}

\begin{figure}[htbp]
\centering
\vspace{-1.5em}
\includegraphics[width=\textwidth]{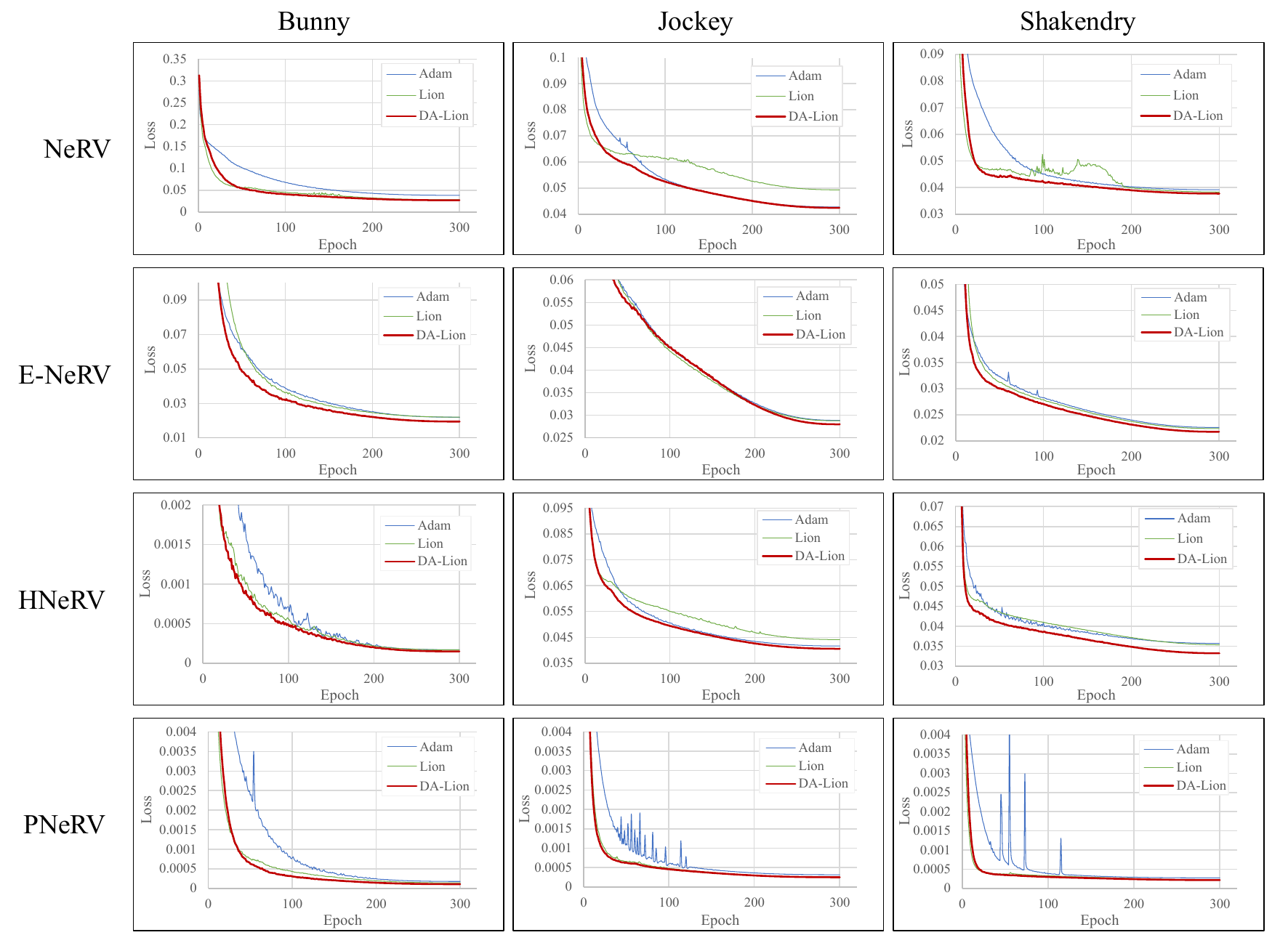}
\caption{Training loss curves of Adam, Lion, and DA-Lion on Bunny, Jockey, and Shakendry across four backbones.}
\label{fig:convergence}
\vspace{-1.0em}
\end{figure}

Fig.~\ref{fig:convergence} compares training loss curves of Adam, Lion, and DA-Lion on Bunny, Jockey, and Shakendry across four backbones.
DA-Lion produces smoother trajectories than Lion and converges faster than Adam while remaining stable in later stages. On Shakendry with NeRV, DA-Lion matches Adam's final loss in roughly 2/3 the epochs.

\begin{figure}[htbp]
\centering
\vspace{-1.5em}
\includegraphics[width=0.95\textwidth]{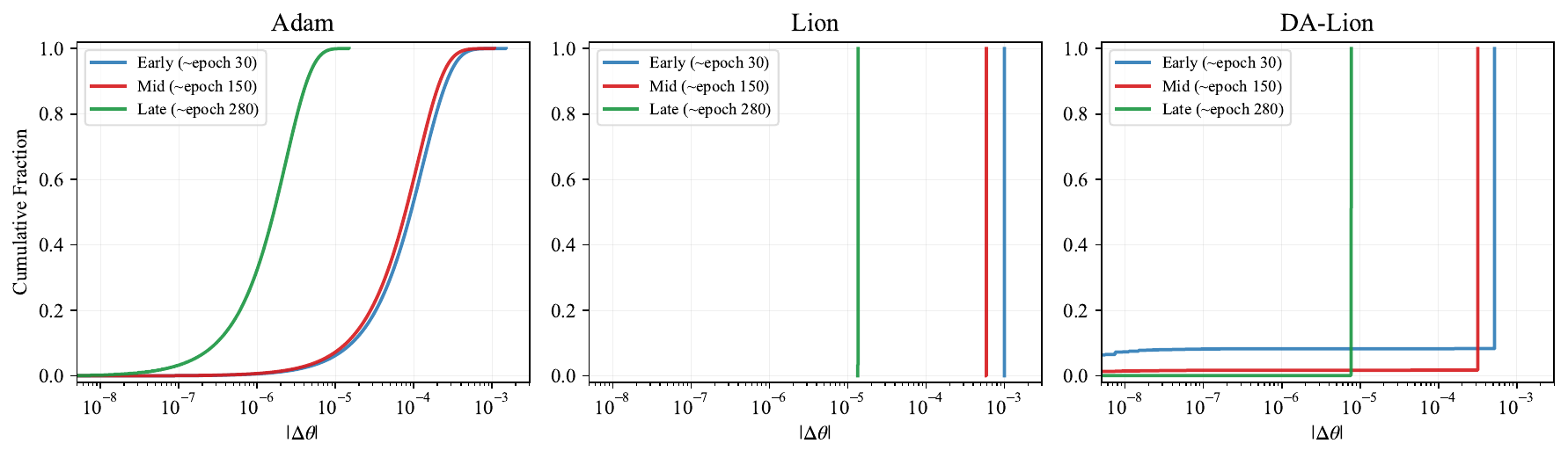}
\caption{Empirical cumulative distribution function (eCDF) of per-parameter $|\Delta\theta|$.}
\label{fig:update_ecdf}
% \vspace{-1.5em}
\end{figure}

To understand the mechanism, we recorded per-parameter update magnitudes $|\Delta\theta|$ during NeRV training on Bunny.
Fig.~\ref{fig:update_ecdf} reports the eCDF of $|\Delta\theta|$ at three training stages.
Adam maintains a smooth distribution spanning orders of magnitude, reflecting gradient-dependent step sizes.
Lion collapses to a degenerate distribution: at the late stage, all parameters share only 3 distinct $|\Delta\theta|$ values clustered around $\eta_t$---the empirical signature of the noise floor predicted in Sec.~\ref{sec:method}.
DA-Lion splits into two groups: coordinates where gradient and momentum disagree (the $1-p_t$ fraction, $\sim$half at the late stage) receive near-zero updates from the momentum-form branch, while the remaining coordinates follow sign updates of magnitude $\alpha_t\eta_t$, producing a steep but non-degenerate eCDF intermediate between Adam's smooth spread and Lion's single-value collapse.

\begin{figure}[htbp]
\centering
\vspace{-1.5em}
\includegraphics[width=0.95\textwidth]{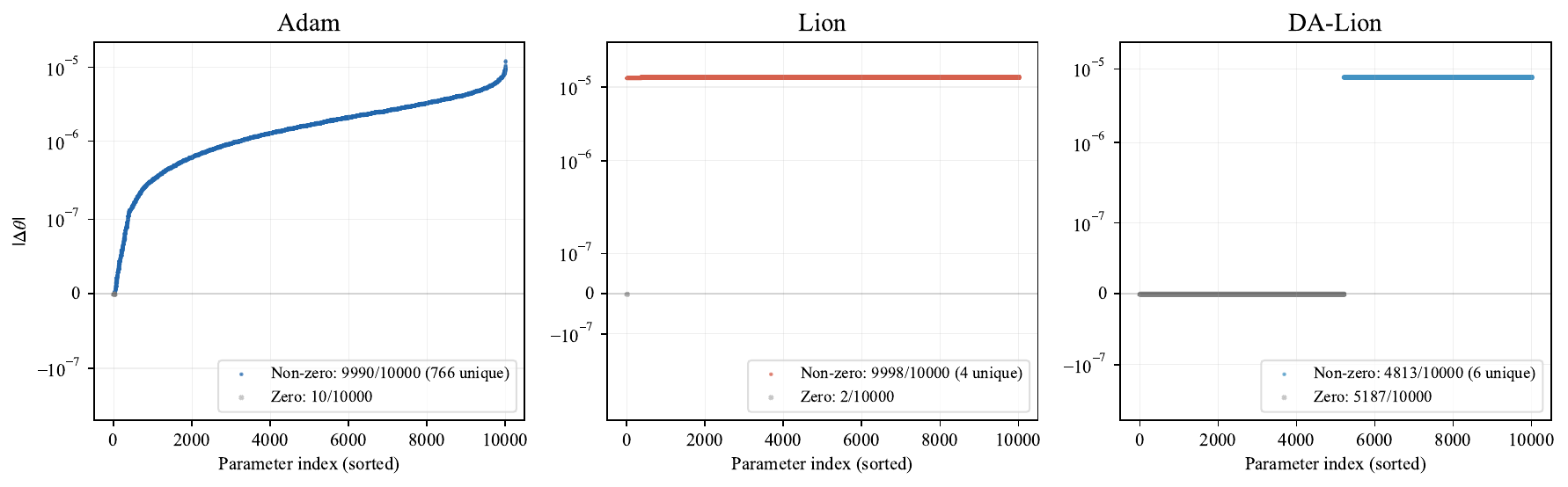}
\caption{Random sample of 10,000 per-parameter $|\Delta\theta|$ at epoch$\sim$280, sorted by value (NeRV on Bunny).}
\label{fig:update_scatter}
\vspace{-1.5em}
\end{figure}

Fig.~\ref{fig:update_scatter} shows 10,000 randomly sampled $|\Delta\theta|$ at the late stage (epoch$\sim$280), sorted by value.
Lion produces a flat line---every coordinate receives an identical update magnitude, confirming the noise floor operates uniformly.
DA-Lion splits coordinates into two groups: disagreeing coordinates at near-zero (gray crosses) and stable-descent coordinates with uniformly scaled sign updates.
Adam yields a smoothly varying curve, confirming that update magnitudes track the per-coordinate gradient scale absent sign discretization.

\subsection{Quantitative results}

Table~\ref{tab:bunny_dif_budget}(a) reports PSNR on Bunny under four model sizes (3.0M--7.5M). DA-Lion achieves the highest average PSNR on all four backbones, with gains over Adam ranging from $+0.39$\,dB (HNeRV) to $+2.06$\,dB (PNeRV). The smaller margin on HNeRV reflects its hybrid encoder-decoder design, which already provides a stronger inductive bias that partially compensates for optimizer-induced instability.
Table~\ref{tab:bunny_dif_budget}(b) further reports results under varying training epochs (300e--1800e), where DA-Lion again achieves the best average PSNR across all four backbones.

\begin{table}[htbp]
\centering
\vspace{-1.5em}
\caption{Quantitative comparison on Bunny (640$\times$1280) under different budgets (PSNR$\uparrow$).}
\label{tab:bunny_dif_budget}
\subfigure[Different model size]{%
\resizebox{\textwidth}{!}{
\begin{tabular}{l*{3}{c}|*{3}{c}|*{3}{c}|*{3}{c}}
\toprule
Backbone& \multicolumn{3}{c}{NeRV\cite{chen2021nerv}} & \multicolumn{3}{c}{E-NeRV\cite{li2022e-nerv}} & \multicolumn{3}{c}{HNeRV\cite{chen2023hnerv}} & \multicolumn{3}{c}{PNeRV\cite{zhao2024pnerv}} \\
\cmidrule(r){1-4} \cmidrule(lr){5-7} \cmidrule(lr){8-10} \cmidrule(lr){11-13}
Size & Adam & Lion & DA-Lion & Adam & Lion & DA-Lion & Adam & Lion & DA-Lion & Adam & Lion & DA-Lion \\
\midrule
3.0M & 32.13 & 33.67 & 34.01 & 34.38 & 34.59 & 35.65 & 38.04 & 38.01 & 38.56 & 38.10 & 39.84 & 39.95 \\
4.5M & 33.60 & 35.05 & 35.46 & 37.03 & 37.26 & 38.11 & 39.29 & 39.42 & 39.88 & 38.81 & 41.44 & 41.44 \\
6.0M & 35.46 & 36.25 & 37.08 & 38.43 & 38.65 & 39.80 & 40.27 & 39.87 & 40.65 & 39.54 & 39.30 & 41.31 \\
7.5M & 36.86 & 37.20 & 38.29 & 39.31 & 39.80 & 40.47 & 40.86 & 40.27 & 40.96 & 38.45 & 39.59 & 40.45 \\ \midrule
Avg. & 34.51 & \cellcolor{red!25}35.54 & \cellcolor{red!50}36.21 & 37.29 & \cellcolor{red!25}37.58 & \cellcolor{red!50}38.51 & \cellcolor{red!25}39.62 & 39.39 & \cellcolor{red!50}40.01 & 38.73 & \cellcolor{red!25}40.04 & \cellcolor{red!50}40.79 \\
\bottomrule
\end{tabular}}}
\\[-0.2em]
\subfigure[Different train epoch]{%
\resizebox{\textwidth}{!}{
\begin{tabular}{l*{3}{c}|*{3}{c}|*{3}{c}|*{3}{c}}
\toprule
Backbone& \multicolumn{3}{c}{NeRV\cite{chen2021nerv}} & \multicolumn{3}{c}{E-NeRV\cite{li2022e-nerv}} & \multicolumn{3}{c}{HNeRV\cite{chen2023hnerv}} & \multicolumn{3}{c}{PNeRV\cite{zhao2024pnerv}} \\
\cmidrule(r){1-4} \cmidrule(lr){5-7} \cmidrule(lr){8-10} \cmidrule(lr){11-13}
Epoch & Adam & Lion & DA-Lion & Adam & Lion & DA-Lion & Adam & Lion & DA-Lion & Adam & Lion & DA-Lion \\
\midrule
$300e$ & 32.13 & 33.67 & 34.01 & 34.38 & 34.59 & 35.65 & 38.04 & 38.01 & 38.56 & 38.10 & 39.84 & 39.95 \\
$600e$ & 33.29 & 34.14 & 34.49 & 35.71 & 35.82 & 36.44 & 39.37 & 38.99 & 39.46 & 40.26 & 39.89 & 40.78 \\
$1200e$ & 34.00 & 34.68 & 34.94 & 36.67 & 36.78 & 37.19 & 40.03 & 39.74 & 40.04 & 41.32 & 40.31 & 41.34 \\
$1800e$ & 34.30 & 35.25 & 34.95 & 37.15 & 37.18 & 37.47 & 40.28 & 40.01 & 40.35 & 41.71 & 40.66 & 41.61 \\ \midrule
Avg. & 33.43 & \cellcolor{red!25}34.44 & \cellcolor{red!50}34.60 & 35.98 & \cellcolor{red!25}36.09 & \cellcolor{red!50}36.69 & \cellcolor{red!25}39.43 & 39.19 & \cellcolor{red!50}39.60 & \cellcolor{red!25}40.35 & 40.18 & \cellcolor{red!50}40.92 \\
\bottomrule
\end{tabular}}}
\vspace{-2.5em}
\end{table}

Table~\ref{tab:uvg_per_seq_3m} extends the comparison to UVG and DAVIS datasets.
On UVG, DA-Lion achieves the best average PSNR for all four backbones.
On DAVIS, DA-Lion yields clear gains over both Adam and Lion on NeRV, E-NeRV, and HNeRV. 
On PNeRV, DA-Lion trails Lion by 0.37 dB on average, underperforming on four of seven DAVIS sequences. 
PNeRV's pyramidal multi-scale features amplify sensitivity to the modulation scale $s_b$ (see Table~\ref{tab:hyper_ablation}(c)), suggesting that backbone-specific tuning of $\alpha_t$ is warranted for such architectures.
Overall, DA-Lion attains the best performance in most configurations across datasets and backbones.

\begin{table}[htbp]
\centering
\vspace{-1.5em}
\caption{Quantitative comparison of video regression on the UVG and DAVIS 2016 dataset (PSNR$\uparrow$).}
\subfigure[UVG]{%
\resizebox{\textwidth}{!}{
\begin{tabular}{l*{3}{c}|*{3}{c}|*{3}{c}|*{3}{c}}
\toprule
Video & \multicolumn{3}{c}{NeRV\cite{chen2021nerv}} & \multicolumn{3}{c}{E-NeRV\cite{li2022e-nerv}} & \multicolumn{3}{c}{HNeRV\cite{chen2023hnerv}} & \multicolumn{3}{c}{PNeRV\cite{zhao2024pnerv}} \\
\cmidrule(r){1-4} \cmidrule(lr){5-7} \cmidrule(lr){8-10} \cmidrule(lr){11-13}
Sequence & Adam & Lion & DA-Lion & Adam & Lion & DA-Lion & Adam & Lion & DA-Lion & Adam & Lion & DA-Lion \\
\midrule
Beauty & 34.06 & 33.37 & 34.15 & 35.10 & 35.06 & 35.19 & 34.84 & 34.67 & 34.92 & 34.94 & 34.94 & 35.20 \\
Bosph & 33.52 & 33.58 & 34.14 & 34.96 & 34.94 & 35.10 & 35.37 & 35.56 & 35.52 & 36.47 & 36.24 & 36.97 \\
Bee & 39.39 & 39.35 & 39.64 & 41.31 & 41.41 & 41.32 & 39.90 & 39.94 & 40.11 & 40.37 & 40.67 & 40.79 \\
Jockey & 31.48 & 29.74 & 31.53 & 30.55 & 30.42 & 30.72 & 32.16 & 31.50 & 32.60 & 35.09 & 35.36 & 36.09 \\
Ready & 24.93 & 23.95 & 25.03 & 25.57 & 25.35 & 25.62 & 25.30 & 25.84 & 25.79 & 29.14 & 29.14 & 29.86 \\
Shakendry & 34.36 & 34.64 & 34.92 & 36.26 & 36.30 & 36.50 & 35.74 & 35.91 & 36.22 & 36.16 & 36.87 & 36.96 \\
Yacht & 27.70 & 27.53 & 28.02 & 28.81 & 28.79 & 28.91 & 29.33 & 29.55 & 29.62 & 31.22 & 30.99 & 31.67 \\ \midrule
Avg. & \cellcolor{red!25}32.21 & 31.74 & \cellcolor{red!50}32.49 & \cellcolor{red!25}33.22 & 33.18 & \cellcolor{red!50}33.34 & 33.23 & \cellcolor{red!25}33.28 & \cellcolor{red!50}33.54 & 34.77 & \cellcolor{red!25}34.89 & \cellcolor{red!50}35.36 \\
\bottomrule
\end{tabular}}}
\\[-0.2em]
\subfigure[DAVIS]{%
\resizebox{\textwidth}{!}{
\begin{tabular}{l*{3}{c}|*{3}{c}|*{3}{c}|*{3}{c}}
\toprule
Video & \multicolumn{3}{c}{NeRV\cite{chen2021nerv}} & \multicolumn{3}{c}{E-NeRV\cite{li2022e-nerv}} & \multicolumn{3}{c}{HNeRV\cite{chen2023hnerv}} & \multicolumn{3}{c}{PNeRV\cite{zhao2024pnerv}} \\
\cmidrule(r){1-4} \cmidrule(lr){5-7} \cmidrule(lr){8-10} \cmidrule(lr){11-13}
Sequence & Adam & Lion & DA-Lion & Adam & Lion & DA-Lion & Adam & Lion & DA-Lion & Adam & Lion & DA-Lion \\
\midrule
Blackswan & 24.37 & 28.01 & 28.14 & 27.85 & 27.87 & 29.41 & 29.73 & 32.24 & 32.39 & 29.64 & 34.33 & 34.54 \\
Bmx-trees & 24.50 & 26.56 & 26.81 & 25.91 & 26.34 & 27.66 & 29.09 & 30.38 & 30.64 & 29.18 & 32.40 & 31.70 \\
Boat & 29.55 & 31.76 & 31.82 & 30.88 & 30.98 & 31.95 & 32.84 & 31.17 & 34.25 & 32.72 & 34.97 & 35.24 \\
Breakdance & 24.45 & 26.84 & 27.17 & 26.54 & 25.62 & 27.97 & 30.71 & 31.74 & 31.85 & 30.00 & 33.44 & 32.82 \\
Surf & 28.45 & 32.78 & 32.84 & 31.79 & 31.63 & 34.18 & 32.40 & 32.68 & 35.25 & 32.75 & 37.42 & 37.48 \\
Swing & 25.81 & 28.54 & 28.68 & 27.67 & 27.79 & 29.52 & 30.70 & 32.60 & 33.42 & 30.14 & 34.84 & 34.10 \\
Goat & 21.77 & 23.87 & 23.85 & 23.83 & 23.72 & 24.69 & 27.41 & 27.66 & 28.03 & 27.19 & 29.36 & 28.27 \\ \midrule
Avg. & 25.56 & \cellcolor{red!25}28.34 & \cellcolor{red!50}28.47 & \cellcolor{red!25}27.78 & 27.71 & \cellcolor{red!50}29.34 & 30.41 & \cellcolor{red!25}31.21 & \cellcolor{red!50}32.26 & 30.23 & \cellcolor{red!50}33.82 & \cellcolor{red!25}33.45 \\
\bottomrule
\end{tabular}}}
\label{tab:uvg_per_seq_3m}
\end{table}

\subsection{Qualitative comparison}
\begin{figure}[htbp]
\centering
\includegraphics[width=0.95\textwidth]{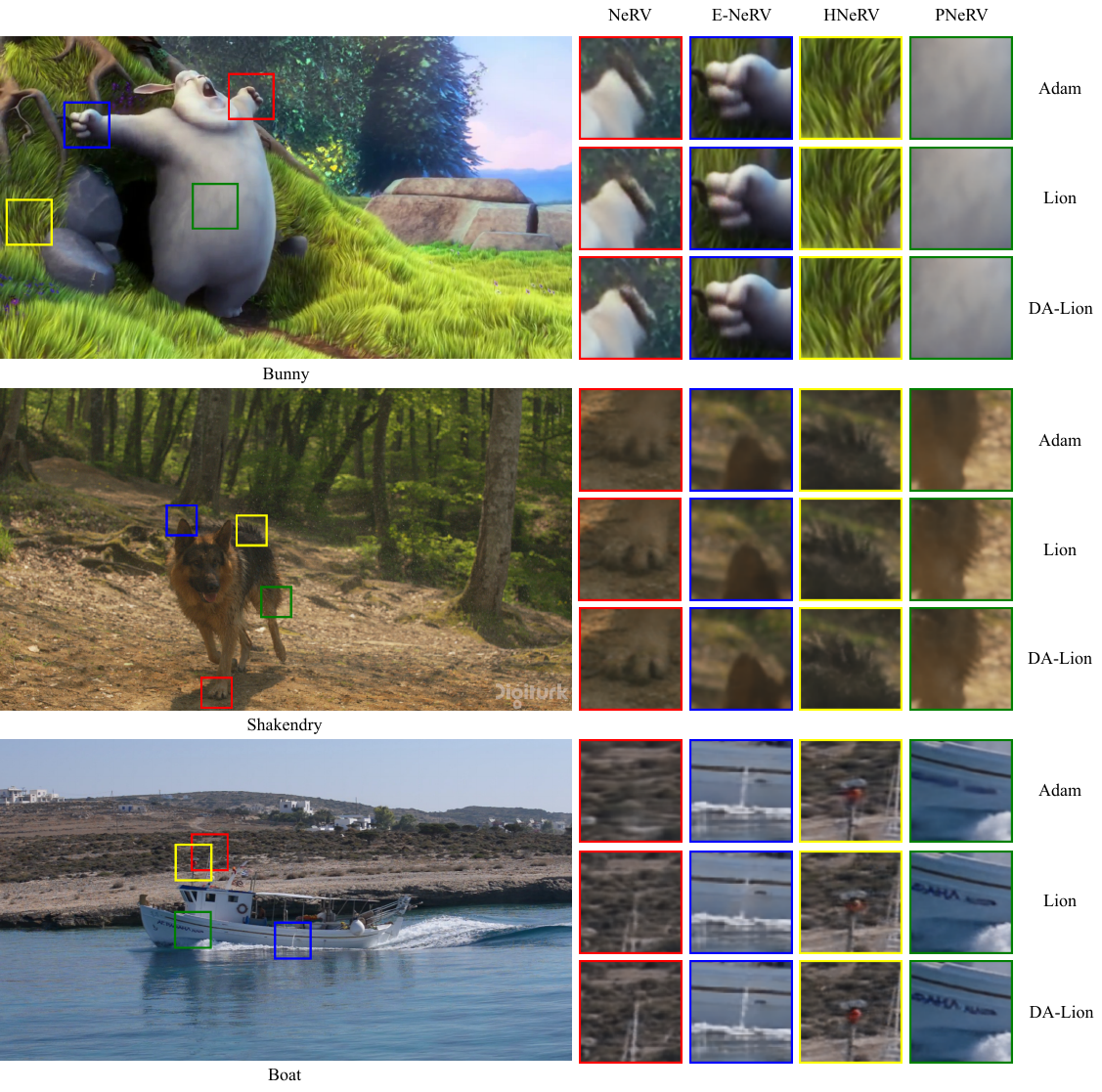}
\caption{Qualitative comparison on Bunny, Shakendry and Boat across four backbones.}
\label{fig:qualitative}
\vspace{-1.5em}
\end{figure}

Fig.~\ref{fig:qualitative} compares reconstructed frames across four backbones on Bunny, Shakendry, and Boat. DA-Lion produces sharper reconstructions with fewer artifacts: Adam over-smooths (hand contours on Bunny, body-part boundaries on Shakendry, text edges on Boat), Lion recovers edges but introduces high-frequency artifacts, and DA-Lion preserves clean edges while suppressing instability across all three sequences.

\subsection{Compression performance}
\begin{figure}[htbp]
\centering
\vspace{-1.5em}
\includegraphics[width=0.95\textwidth]{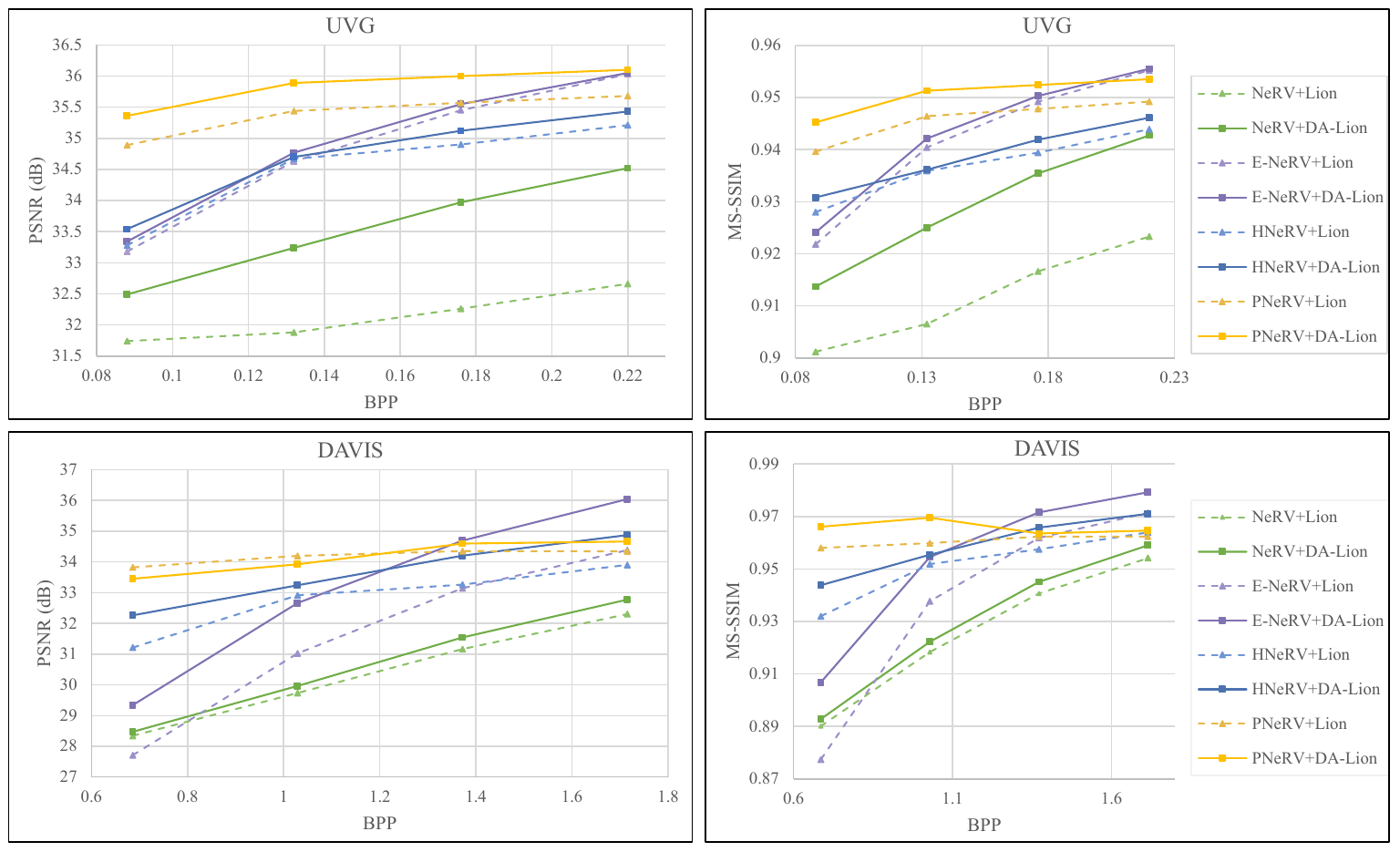}
\caption{Compression performance comparisons in UVG and DAVIS datasets.}
\label{fig:rd}
\vspace{-1.5em}
\end{figure}

Since DA-Lion modifies only the parameter update rule, we adopt each backbone's standard compression pipeline without any modification.
Fig.~\ref{fig:rd} reports rate--distortion curves on UVG and DAVIS.
On UVG, DA-Lion improves over Lion across all backbones, with clear gains on NeRV and PNeRV; Lion underperforms Adam at several bitrates on NeRV, whereas DA-Lion surpasses both.
On DAVIS, improvements are more modest but uniformly positive---the dataset's challenging content (fast motion, occlusions, complex scene transitions) narrows optimizer performance differences.

\subsection{Ablations and complexity}
Table~\ref{tab:ablation} ablates the optimizer design on NeRV across two DAVIS sequences. Results are reported as mean$\pm$standard deviation over five random seeds (1, 42, 123, 456, 789) with identical hyperparameters.
Lion with a halved learning rate (Lion-LR, lr${=}0.0005$) improves over the default Lion (lr${=}0.001$) by $+0.40$--$+0.53$\,dB, confirming that effective step-size calibration contributes meaningfully to the gains observed in alignment-based variants.
Mask-Lion and Consis-Lion both apply the direction-consistency criterion but differ in the treatment of disagreeing coordinates: Mask-Lion masks them to zero while Consis-Lion switches to a momentum-form update.
Their performance is nearly identical across both sequences ($\Delta{<}0.05$\,dB), suggesting that the gradient--momentum alignment signal itself, rather than the specific disagreement strategy, is the primary source of improvement over vanilla Lion.
DA-Lion further incorporates learning-rate-aware magnitude modulation and achieves the best overall PSNR and MS-SSIM on both videos.

\begin{table}[htbp]
\centering
\vspace{-2.5em}
\caption{Ablation on optimizer design (NeRV on Bmx-trees and Breakdance).}
\begin{tabular}{l|lcccc}
\toprule
Video &Optimizer & Consistency & Modulation & PSNR$\uparrow$ & MS-SSIM$\uparrow$ \\
\midrule
\multirow{6}{*}{Bmx-trees} &Adam &  &  & 28.12$\pm$0.06 & 0.9124$\pm$0.0012 \\
 &Lion &  &  & 28.25$\pm$0.16 & 0.9143$\pm$0.0041 \\
 &Lion-LR &  &  & 28.78$\pm$0.12 & 0.9228$\pm$0.0015 \\
 &Mask-Lion & \Checkmark &  & 29.12$\pm$0.11 & 0.9288$\pm$0.0033 \\
 &Consis-Lion & \Checkmark &  & 29.16$\pm$0.14 & 0.9296$\pm$0.0034 \\
 &DA-Lion & \Checkmark & \Checkmark & \cellcolor{red!50}30.09$\pm$0.07 & \cellcolor{red!50}0.9438$\pm$0.0010 \\
\midrule
\multirow{6}{*}{Breakdance} &Adam &  &  & 26.58$\pm$0.07 & 0.9562$\pm$0.0005 \\
 &Lion &  &  & 26.74$\pm$0.05 & 0.9596$\pm$0.0005 \\
 &Lion-LR &  &  & 27.14$\pm$0.09 & 0.9630$\pm$0.0006 \\
 &Mask-Lion & \Checkmark &  & 27.48$\pm$0.04 & 0.9659$\pm$0.0005 \\
 &Consis-Lion & \Checkmark &  & 27.53$\pm$0.07 & 0.9664$\pm$0.0006 \\
 &DA-Lion & \Checkmark & \Checkmark & \cellcolor{red!50}28.00$\pm$0.04 & \cellcolor{red!50}0.9704$\pm$0.0003 \\
\bottomrule
\end{tabular}
\label{tab:ablation}
\vspace{-1.5em}
\end{table}

Table~\ref{tab:hyper_ablation} examines DA-Lion's robustness to key hyperparameters.
$\beta_1$ has negligible effect across $\{0.80,0.85,0.90,0.95\}$; decreasing $\beta_2$, however, degrades PSNR substantially, confirming that long-horizon momentum estimates are essential under deterministic NVR gradients.
The modulation scale $s_b$ is nearly flat on NeRV for $s_b\in[0.8,1.5]$, but larger scales on HNeRV cause significant degradation due to amplified overshoot from content-conditioned features.
We use $s_b{=}1.0$ and inherit Lion's $(\beta_1,\beta_2){=}(0.9,0.99)$ as robust defaults.

\begin{table}[htbp]
\centering

\caption{Hyperparameter ablation of DA-Lion on Bunny (3.0\,M, 300 epochs, PSNR/MS-SSIM). Red rows mark the default setting adopted in all other experiments.}
\label{tab:hyper_ablation}
\subfigure[$(\beta_1,\beta_2)$ on NeRV]{
\begin{tabular}{cc|c}
\toprule
$\beta_1$ & $\beta_2$ & PSNR/MS-SSIM \\
\midrule
0.80 & 0.99 & 33.96/0.9742 \\
0.85 & 0.99 & \cellcolor{red!25}33.98/0.9743 \\
0.95 & 0.99 & 33.96/0.9742 \\
0.90 & 0.95 & 33.40/0.9705 \\
0.90 & 0.97 & 33.57/0.9718 \\
0.90 & 0.99 & \cellcolor{red!50}34.01/0.9743 \\
\bottomrule
\end{tabular}}
\hfill
\subfigure[$s_b$ on NeRV]{
\begin{tabular}{c|c}
\toprule
$s_b$ & PSNR/MS-SSIM \\
\midrule
0.5 & 33.93/0.9739 \\
0.8 & 34.00/0.9743 \\
1.0 & \cellcolor{red!25}34.01/0.9744 \\
1.5 & \cellcolor{red!50}34.02/0.9744 \\
\bottomrule
\end{tabular}}
\hfill
\subfigure[$s_b$ on HNeRV]{
\begin{tabular}{c|c}
\toprule
$s_b$ & PSNR/MS-SSIM \\
\midrule
1.0 & \cellcolor{red!50}38.56/0.9892 \\
1.5 & \cellcolor{red!25}37.84/0.9873 \\
2.0 & 36.98/0.9844 \\
3.0 & 35.10/0.9760 \\
\bottomrule
\end{tabular}}
\end{table}

Table~\ref{tab:complexity} confirms that DA-Lion preserves model size, FLOPs, and GPU memory, with only a minor training-time overhead ($+2.96\%$ vs.~Adam). We also report training stability (TS), measured as the standard deviation of the first-order difference of the training loss across the final 50 epochs; a lower TS indicates smoother convergence with less oscillation.

\begin{table}[htbp]
\centering
\caption{Complexity and efficiency comparison (NeRV 3.0M on Bunny, mean$\pm$std over 5 seeds).}
\begin{tabular}{l|cccc|ccc}
\toprule
Optimizer  & GPU  & Enc. Time (s) & PSNR & TS & $\Delta$Enc. Time$\downarrow$ & $\Delta$PSNR$\uparrow$ &$\Delta$TS$\downarrow$ \\
\midrule
Adam    & 2.47G  & 1217 & 32.13$\pm$0.08 & 3.4$\pm$0.3e-3 & --- & --- & --- \\
Lion    & 2.47G  & 1224 & 33.67$\pm$0.15 & 4.4$\pm$0.5e-3 & \cellcolor{red!50}+0.58\% & +1.54 & +29.41\% \\
DA-Lion & 2.47G  & 1253 & 34.01$\pm$0.06 & 3.7$\pm$0.4e-3 & +2.96\% & \cellcolor{red!50}+1.88 & \cellcolor{red!50}+8.82\% \\
\bottomrule
\end{tabular}
\label{tab:complexity}
\end{table}

\section{Conclusion}
This paper proposes Direction-Aware Lion (DA-Lion), an optimizer tailored for neural video representation (NVR), where training is instance-specific signal fitting and optimization dynamics directly affect reconstruction quality under a fixed budget.
Building on Lion, DA-Lion introduces two lightweight mechanisms: (i) a direction-consistency criterion that switches between a sign update and a momentum-form update based on the alignment between the current gradient and historical momentum, and (ii) a learning-rate-aware magnitude modulation that stabilizes sign step sizes across training phases.
Experiments show that DA-Lion improves convergence stability and boosts PSNR/MS-SSIM.
Since DA-Lion changes only the update rule, it preserves model architecture, parameter count, and FLOPs, while achieving smoother convergence  with only a minor training-time overhead.
This work demonstrates that adapting optimizer design to the deterministic instance-fitting characteristics of INR can yield practical gains, and we hope it encourages further exploration of optimization algorithms tailored for neural representation learning.

\textbf{Acknowledgments.} This work was supported by the National Natural Science Foundation of China (Grant No.62571160), Engineering Technology R\&D Center of Guangdong Provincial Universities (2024GCZX004), and the Pengcheng Laboratory.

\bibliographystyle{splncs04}
\bibliography{references}

\end{document}